\documentclass[showpacs,preprintnumbers,amsmath,amssymb]{revtex4}

\usepackage{epsfig}
\usepackage{graphicx}% Include figure files
\usepackage{dcolumn}% Align table columns on decimal point
\usepackage{bm}% bold math

\begin{document}

\title{Causal interactions and delays in a neuronal ensemble}

\author{Daniele Marinazzo$^{1,2,3}$, Mario Pellicoro$^{1,2,3}$, Sebastiano Stramaglia$^{1,2,3}$}

 \affiliation{
$^1$TIRES-Center of
Innovative Technologies for Signal Detection and Processing,\\
Universit\`a di Bari, Italy\\
$^2$Dipartimento Interateneo di Fisica, Bari, Italy \\
$^3$Istituto Nazionale di Fisica Nucleare, Sezione di Bari, Italy  }

\date{\today}

\begin{abstract}
We analyze a neural system which mimics a sensorial cortex, with
different input characteristics, in presence of transmission delays.
We propose a new measure to characterize collective behavior, based
on the nonlinear extension of the concept of Granger causality, and
an interpretation is given of the variation of the percentage of the
causally relevant interactions with transmission delays.


\end{abstract}

\maketitle

\section{Introduction}
Orientation selectivity and perception are connected with the
collective behavior in neural ensembles \cite{BenYishai1995},
\cite{Moore2001}. Previous work has shown that networks of
excitatory and inhibitory leaky integrate-and-fire neurons tend to
oscillate under some general conditions \cite{Borgers2003}, provided
that the excitatory neurons receive a sufficiently large input,
while other studies have shown how oscillatory activity depends on
the spatiotemporal properties of the external input
\cite{Doiron2004}. Due to the finite-velocity propagation of action
potentials and to the spike generation dynamics, the presence of
delays is physiological in networks of neurons. Sometimes delays are
seen as an annoying presence and are thus neglected. In other
studies is shown that they can give rise to a wide gamma of
behaviors \cite{Roxin2005}; also groups of neurons with reproducible
time-locked but not synchronous firing patterns have been
individuated \cite{Izhikevich2006}. Furthermore, the role of
synchronization is controversial. Undoubtedly, when this phenomenon
is limited to a small time interval, it is the index of something
going on. On the other hand, when facing a fully synchronized
network, is difficult to extract any kind of information when all
the neurons behave as a single one. We try then to get a better
insight on this issue starting from the idea that the essence of
collective behavior is the presence of causal interactions. We thus
want to individuate the causally relevant relationships between the
neurons in the network. The notion of Granger causality
\cite{Granger1969} between two time series examines if the
prediction of one series could be improved by incorporating
information of the other. In particular, if the prediction error of
the first time series is reduced by including measurements from the
second time series, then the second time series is said to have a
causal influence on the first one. The interactions between
individual neurons in a network are nonlinear. We thus propose a
radial basis function approach to nonlinear Granger causality
\cite{Marinazzopre2006}, and show how these causal influences are
related to the input signal and to the internal delays.

\section{The model}

Our network is a basic model of a mammal cortex, similar to the one
in \cite{Izhikevich2006}, and consists of $N_{E}=400$ excitatory
neurons and $N_{I}=100$ inhibitory neurons. Each excitatory neuron
is connected to $50$ random neurons, both excitatory and inhibitory,
while each inhibitory neuron is connected to $50$ excitatory random
neurons.

The membrane potential of a LIF neuron satisfies:

\begin{equation}\label{lifmemb}
\frac{dV(t)}{dt}=-(V(t)-V_{r})+I(t),
\end{equation}

where the membrane resistance is normalized to one. Every time, when
the potential of the neuron reaches the threshold value $V_{th}$, a
spike is fired. This resets the potential to the rest potential
$V_r$ and remains bound to this value for an absolute refractory
period $\tau_{ref}$.  Each inhibitory neuron $j$ ($j=1 \ldots
N_{I}$) receives an input $I_{j}(t)$:

\begin{equation}\label{inputinh}
I_{j}(t)=\mu + \eta_{j}(t) + ST_{E,j}(t),
\end{equation}

which consists of a constant base current $\mu$, internal Gaussian
white noise $\eta$ with intensity $D$, and where $ST_{E,j}(t)$ is
the sum of the post-synaptic potentials (PSPs) of the afferent
excitatory neurons.

On the other end, each excitatory neuron $i$ ($i=1 \ldots N_{E}$)
receives an input $I_{i}(t)$:
\begin{equation}\label{inputexc}
I_{i}(t)=\mu + \eta_{i}(t) +ST_{E,i}(t)+ ST_{I,i}(t)+
\sigma[\sqrt{1-c}\xi_{i}(t)+\sqrt{c}\xi_{G}(t)],
\end{equation}

with the same internal current $\mu$ and noise $\eta$ as in the
previous equation, and where $ST_{E,i}(t)$ and $ST_{I,i}(t)$ are the
sums of the PSPs of the afferent excitatory and inhibitory neurons,
respectively. Furthermore we have an additional term
$s_{j}(t)=\sigma[\sqrt{1-c}\xi_{j}(t)+\sqrt{c}\xi_{G}(t)]$, where
$\xi_{j}(t)$ and $\xi_{G}(t)$ are both Gaussian white noise with
zero mean and unit power. This term mimics an external stimulus.
Varying $c$ increases or decreases the degree of spatial correlation
of the external stimuli, while the total input power to each neuron
remains constant. In our model we use dynamical depressing synapses,
such that PSPs are delivered through synapses whose effective
strength is given by the following equations \cite{Tsodyks1997}:

\begin{eqnarray}
\begin{array}{l}
\frac{dx}{dt}=\frac{z}{\tau_{rec}}-U x s^{P}(t),\\
\frac{dy}{dt}=-\frac{y}{\tau_{in}}-U x s^{P}(t),\\
\frac{dz}{dt}=\frac{y}{\tau_{in}}-\frac{z}{\tau_{rec}},\\
\label{dyn}\
\end{array}
\end{eqnarray}

where x, y and z are the fraction of synaptic resources in the
recovered, active and inactive state, respectively, and $s^{P}(t)$
is the sum of all the presynaptic activity at time t. Without any
spike input all neurotransmitter is recovered and the fraction of
available neurotransmitter is one: $x(t)=1$. After each spike
arriving at the synapse, a fraction U of the available (recovered)
neurotransmitter is released. The fraction y of active
neurotransmitter is then inactivated into the inactive state z.
$\tau_{in}$ is the time constant of the inactivation process and
$\tau_{rec}$ is the recovery time constant for conversion of the
inactive to the active state.

Furthermore we have inserted time delays in the connections. We
performed different simulations, where the delays in excitatory
(inhibitory) connections were randomly chosen between zero and a
maximum value $\tau_{E,max}$ ($\tau_{I,max}$). This maximum value
was varied from zero to $40$ ms, with a step of $2.5$ ms. This
values take into account of the measurements performed in mammal
cortex \cite{Swadlow1992}.

We used the following parameter values in the model simulations :
$V_r$=0, $V_{th}$=1, $\tau_{ref}$=3 ms, $\tau_{m}$=10 ms, base
current $\mu$=0.5, intensity of the internal Gaussian white noise
$D$=0.08, $\sigma$=0.4 (or 0 in case of absence of external input),
$\tau_{in}$=3 ms, $\tau_{rec}$=800 ms, $U(e)$=0.5. All simulations
were integrated using an Euler integration scheme with a time step
of 0.1 ms.

\section{Granger Causality}

Let $\{\bar{x}_i\}_{i=1,.,N}$ and $\{\bar{y}_i\}_{i=1,.,N}$ be two
time series of $N$ simultaneously measured quantities. In the
following we will assume that time series are stationary. We aim at
quantifying {\it how much} $\bar{y}$ {\it is cause of} $\bar{x}$.
For $k=1$ to $M$ (where $M=N-m$, $m$ being the order of the model),
we denote $x^k=\bar{x}_{k+m}$, ${\bf X}^k=(\bar{x}_{k+m-1},
\bar{x}_{k+m-2},...,\bar{x}_{k})$, ${\bf Y}^k=(\bar{y}_{k+m-1},
\bar{y}_{k+m-2},...,\bar{y}_{k})$ and we treat these quantities as
$M$ realizations of the stochastic variables ($x$, ${\bf X}$, ${\bf
Y}$) \footnote{the series are normalized to zero mean and unit
variance}. Let us now consider the general nonlinear model
\begin{eqnarray}
\begin{array}{l}
x=w_0 + {\bf w_{1}}\cdot {\bf \textsl{F}}\left({\bf X}\right)+{\bf w_{2}}\cdot {\bf \textsl{S}}\left({\bf Y}\right)+{\bf w_{3}}\cdot {\bf \textsl{K}}\left({\bf X},{\bf Y}\right),\\
%y={\bf w_{21}}\cdot {\bf \textsl{F}}\left({\bf X}\right)+{\bf w_{22}}\cdot {\bf \textsl{S}}\left({\bf
%Y}\right)+{\bf w_{23}}\cdot {\bf \textsl{K}}\left({\bf X},{\bf Y}\right),
\label{mod-non}
\end{array}
\end{eqnarray}
where $w_0$ is the bias term, $\{\bf w\}$ are real vectors of free
parameters, ${\bf
\textsl{F}}=\left(\varphi_1,...,\varphi_{n_x}\right)$ are $n_{x}$
given nonlinear real functions of $m$ variables, ${\bf
\textsl{S}}=\left(\psi_1,...,\psi_{n_y}\right)$ are $n_y$ other real
functions of $m$ variables, and ${\bf
\textsl{K}}=\left(\xi_1,...,\xi_{n_{xy}}\right)$ are $n_{xy}$
functions of $2m$ variables. Parameters $w_0$ and $\{\bf w\}$ must
be fixed to minimize the prediction error (we assume $M \gg
1+n_{x}+n_{y}+n_{xy}$):
\begin{eqnarray}
\begin{array}{l}
\epsilon_{xy}={1\over M}\sum_{k=1}^M \left(x^k-w_0-{\bf w_{1}}\cdot {\bf \textsl{F}}\left({\bf X}^k\right)-{\bf w_{2}}\cdot {\bf \textsl{S}}\left({\bf Y}^k\right)+{\bf w_{3}}\cdot {\bf \textsl{K}}\left({\bf X}^k,{\bf Y}^k\right)\right)^2.\\
\end{array}
\end{eqnarray}
We also consider the model:
\begin{eqnarray}
\begin{array}{l}
x=v_0+{\bf v_{1}}\cdot {\bf \textsl{F}}\left({\bf X}\right),\\
\end{array}
\label{mmod}
\end{eqnarray}
and the corresponding prediction error $\epsilon_{x}$. If the
prediction of $\bar{x}$ improves by incorporating the past values of
$\{\bar{y}_i\}$, i.e. $\epsilon_{xy}$ is smaller than
$\epsilon_{x}$, then $y$ is said to have a causal influence on $x$.
We must require that, if ${\bf Y}$ is statistically independent of
$x$ and ${\bf X}$,  then $\epsilon_{xy}=\epsilon_{x}$ at least for
$M\to \infty$. For a detailed discussion on how this condition is
achieved in the present case, see \cite{Marinazzopre2006} and
\cite{Ancona2006}.

We choose the functions ${\bf \textsl{F}}$, ${\bf \textsl{S}}$ and
${\bf \textsl{K}}$, in model (\ref{mod-non}), in the frame of Radial
Basis Function (RBF) methods. We fix $n_x =n_y =n_{xy}=n\ll M$: $n$
centers $\{{\bf \tilde{X}}^\rho,{\bf \tilde{Y}}^\rho\}_{\rho=1}^n$,
in the space of $({\bf X},{\bf Y})$ vectors, are determined by a
clustering procedure applied to data $\{({\bf X}^k,{\bf
Y}^k)\}_{k=1}^M$. To find prototypes we use fuzzy c-means, a well
known algorithm which introduces {\it fuzzy} memberships to
clusters, so that a point may belong to several clusters with some
degree in the range [0,1]: in calculating the center of a cluster
the coordinates of each instance are weighted by the value of the
membership function. We then make the following choice for
$\rho=1,\ldots,n$:
\begin{eqnarray}
\begin{array}{ll}
\varphi_\rho \left({\bf X}\right)&=\exp\left({-\|{\bf X}-{\bf \tilde{X}}^\rho\|^2/2\sigma^2}\right),\\
\psi_\rho \left({\bf Y}\right)&=\exp\left({-\|{\bf Y}-{\bf \tilde{Y}}^\rho\|^2 /2\sigma^2}\right),\\
\xi_\rho \left({\bf X},{\bf Y}\right)&=\varphi_\rho \left({\bf
X}\right)\psi_\rho \left({\bf Y}\right),
%\exp-\left({\|{\bf X}-{\bf \tilde{X}}^\rho\|^2
%+\|{\bf Y}-{\bf \tilde{Y}}^\rho\|^2}\right)/ 2\sigma^2&=
\end{array}
\label{eq-rbf}
\end{eqnarray}
$\sigma$ being a fixed parameter, whose order of magnitude is the
average spacing between the centers. The RBF model here proposed can
approximate any function of $\textbf{X}$ and $\textbf{Y}$. We
conclude this section stressing that, according to our experience,
the proposed method is insensitive to details of the clustering
procedure used to find prototypes, provided that $n$ is at least two
orders of magnitude smaller than $M$.

\section{Results and Discussion}

The simulation is run for $30$ seconds of model time, such that the
network is allowed to stabilize itself, and then for further $60$
seconds ($60000$ points). We extracted then $12$ time series of
length $5000$ points, from the PSP of any presynaptic neuron, and
from the membrane potential of the corresponding postsynaptic
neuron. Doing this we take also into account the effects of the
external input on the subthreshold behavior of the neurons
\cite{Gerstner1996}. The series were checked for covariance
stationarity with a Dickey-Fuller test ($p<0.01$). The Granger
causality algorithm was then applied. We identified statistically
relevant interactions performing an F-test (Levene test) of the null
hypothesis that the error on the prediction of one series is not
decreased when information on the other series is added to the
model. This analysis was repeated for every possible delay time in
both excitatory and inhibitory connections, up to $\tau_{E,max}$ and
$\tau_{I,max}$, and the results were averaged over the $12$ trials.

\begin{figure}[ht!]
  \includegraphics[height=.3\textheight]{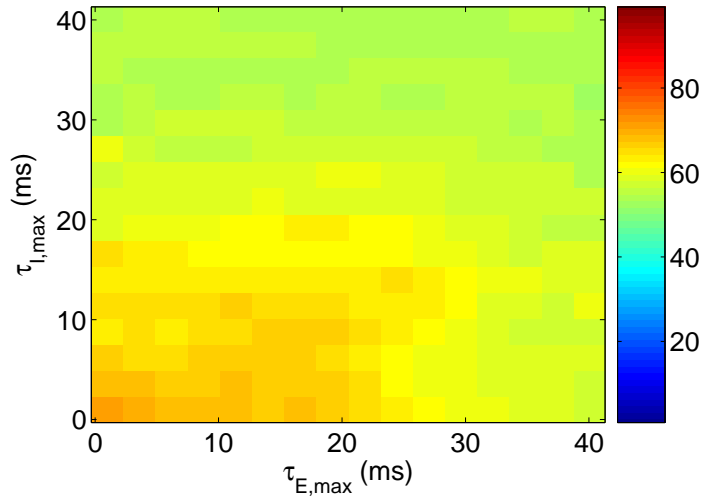}
  \caption{Percentage of causally relevant interactions as a function of the delays
   of excitatory connections (horizontal axis), and inhibitory connections (vertical axis),
   in absence of external input}\label{zerofig}
\end{figure}

In absence of external input ($\sigma$=0), the percentage of
statistically relevant causal interactions decreases slowly but
uniformly with $\tau_{E,max}$ and $\tau_{I,max}$ (Figure
\ref{zerofig}).

\begin{figure}[ht!]
  \includegraphics[height=.3\textheight]{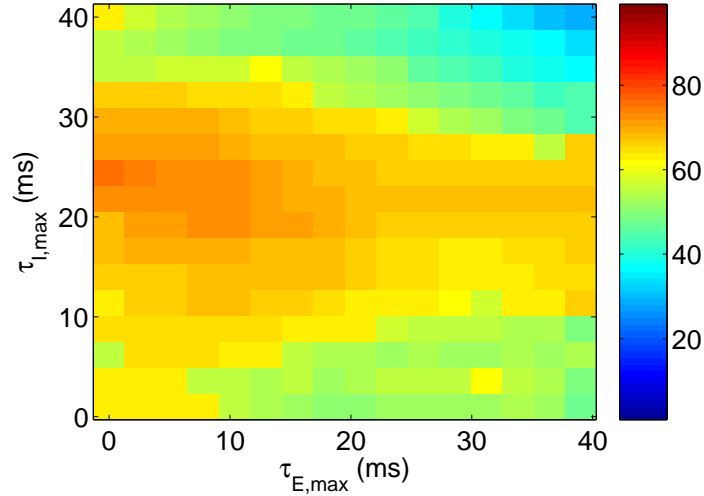}
  \caption{Percentage of causally relevant interactions as a function of the delays
   of excitatory connections (horizontal axis), and inhibitory connections (vertical axis),
   in the case of spatially uncorrelated input to the excitatory
   neurons}\label{uncorfig}
\end{figure}

In the presence of spatially uncorrelated external input (c=0),
there is some structure in the percentage of causal interactions.
The major differences are observed varying the delays in the
inhibitory connections, with a maximum between $20$ and $25$
milliseconds. This region characterized by an increased amount of
relevant interactions becomes narrower as the maximum delay in
excitatory connections increases (Figure \ref{uncorfig}).

\begin{figure}[ht!]
  \includegraphics[height=.3\textheight]{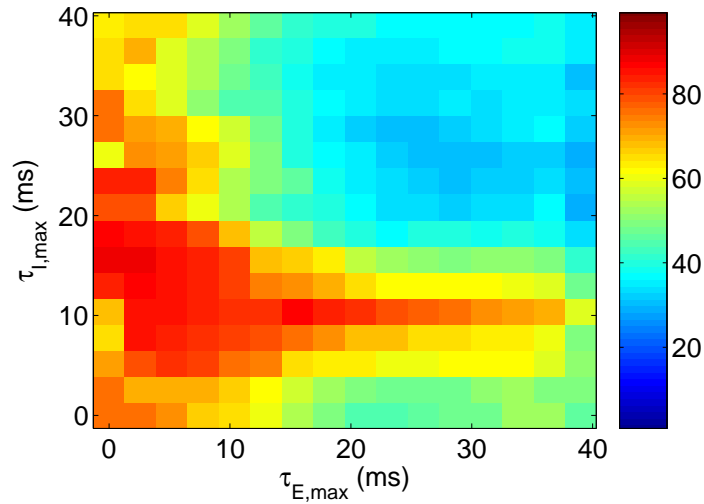}
  \caption{Percentage of causally relevant interactions as a function of the delays
   of excitatory connections (horizontal axis), and inhibitory connections (vertical axis),
   in presence of spatially correlated input to the excitatory neurons}\label{corfig}
\end{figure}

When the external input is spatially correlated (c=1), the structure
becomes much more pronounced, this time with a maximum around
$\tau_{I,max}$ =15 ms for $\tau_{E,max}$= 0, which becomes narrower
and drifts down to $\tau_{I,max}$ = 10 ms for increasing values of
$\tau_{E,max}$ (Figure \ref{corfig}).

In order to explain the structures that appear in presence of
external input, some remarks are in order. A delayed inhibitory
feedback is necessary in order to discriminate an external input
\cite{Doiron2004}, and this explains the lack of causal interactions
in the absence of delays in the inhibitory connections. Furthermore,
if too many inhibitory spikes reach an excitatory neuron at the same
time, the overall inhibitory effect is depressed. On the other hand,
when the delays in the inhibitory connections are scattered along a
too wide interval, the feedback effect is inactivated. As it
concerns the delays in the excitatory connections, it is worth to
recall that a neuron behaves as a coincidence detector when its time
constant is small, changing into an integrator when the time
constant increases. Since the external input is continuous, the
information is constantly carried to the inhibitory neurons, and
this makes the number of causally relevant interaction less
sensitive to the value of $\tau_{E,max}$. Though, when the input is
spatially correlated, the timing is more important if we don't want
to lose the important information that all excitatory neurons
receive the same input at the same time. We observe that the
narrowing of the region with highest percentage of causally relevant
interactions becomes more critical as $\tau_{E,max}$ increases,
remaining optimal only in a small region around a value of
$\tau_{I,max}$ for which the excitatory neurons are still able to
resolve the individual inhibitory spikes. This preferred value in
inhibitory delays can be useful in choosing the optimal window
length in the case of spike-time dependent
plasticity\cite{Song2000}.

\section{Conclusions}
We have introduced the concept of causality, which can give a better
insight on the collective behavior in neural systems. We have shown
how to extend the original definition of causality to nonlinear
systems. We have built a basic model of a sensory cortex and
performed a quantitative analysis of the causally relevant
interactions for different characteristics of the external input.

%%%%%%%%%%%%%%%%%%%%%%%%%%%%%%%%%%%%%%%%%%%%%%%%
%% BACKMATTER
%%%%%%%%%%%%%%%%%%%%%%%%%%%%%%%%%%%%%%%%%%%%%%%%

\begin{center}\vskip 0.4 cm\par\noindent{\bf ACKNOWLEDGEMENTS}\end{center}

We thank Nicola Ancona and Stan Gielen for helpful discussions.

\bibliography{causality_neurons}

\begin{thebibliography}{10}

\bibitem{BenYishai1995}
R~Ben-Yishai, R~L Bar-Or, and H~Sompolinsky.
\newblock Theory of orientation tuning in visual cortex.
\newblock {\em Proc Natl Acad Sci U S A}, 92(9):3844--3848, Apr 1995.

\bibitem{Moore2001}
D~R Moore, J~W Schnupp, and A~J King.
\newblock Coding the temporal structure of sounds in auditory cortex.
\newblock {\em Nat Neurosci}, 4(11):1055--1056, Nov 2001.

\bibitem{Borgers2003}
C~B{\"o}rgers and N~Kopell.
\newblock Synchronization in networks of excitatory and inhibitory neurons with
  sparse, random connectivity.
\newblock {\em Neural Comput}, 15(3):509--538, Mar 2003.

\bibitem{Doiron2004}
Brent Doiron, Benjamin Lindner, Andre Longtin, Leonard Maler, and Joseph
  Bastian.
\newblock Oscillatory activity in electrosensory neurons increases with the
  spatial correlation of the stochastic input stimulus.
\newblock {\em Physical Review Letters}, 93(4):048101, 2004.

\bibitem{Roxin2005}
Alex Roxin, Nicolas Brunel, and David Hansel.
\newblock Role of delays in shaping spatiotemporal dynamics of neuronal
  activity in large networks.
\newblock {\em Physical Review Letters}, 94(23):238103, 2005.

\bibitem{Izhikevich2006}
E~M Izhikevich.
\newblock Polychronization: computation with spikes.
\newblock {\em Neural Comput}, 18(2):245--282, Feb 2006.

\bibitem{Granger1969}
C~W~J Granger.
\newblock Investigating causal relations by econometric models and
  cross-spectral methods.
\newblock {\em Econometrica}, 37(3):424--38, July 1969.

\bibitem{Marinazzopre2006}
Daniele Marinazzo, Mario Pellicoro, and Sebastiano Stramaglia.
\newblock Nonlinear parametric model for granger causality of time series.
\newblock {\em Physical Review E (Statistical, Nonlinear, and Soft Matter
  Physics)}, 73(6):066216, 2006.

\bibitem{Tsodyks1997}
M~V Tsodyks and H~Markram.
\newblock The neural code between neocortical pyramidal neurons depends on
  neurotransmitter release probability.
\newblock {\em Proc Natl Acad Sci U S A}, 94(2):719--723, Jan 1997.

\bibitem{Swadlow1992}
H~A Swadlow.
\newblock Monitoring the excitability of neocortical efferent neurons to direct
  activation by extracellular current pulses.
\newblock {\em J Neurophysiol}, 68(2):605--619, Aug 1992.

\bibitem{Ancona2006}
N~Ancona and S~Stramaglia.
\newblock An invariance property of predictors in kernel-induced hypothesis
  spaces.
\newblock {\em Neural Comput}, 18(4):749--759, Apr 2006.

\bibitem{Gerstner1996}
W~Gerstner, R~Kempter, J~L van Hemmen, and H~Wagner.
\newblock A neuronal learning rule for sub-millisecond temporal coding.
\newblock {\em Nature}, 383(6595):76--81, Sep 1996.

\bibitem{Song2000}
S.~Song, K.~Miller, and L.~Abbott.
\newblock Competitive hebbian learning through spiketime -dependent synaptic
  plasticity.
\newblock {\em Nat. Neurosci.}, 3:919--926, 2000.

\end{thebibliography}

\end{document}